\documentclass[twocolumn,showpacs,preprintnumbers,amsmath,amssymb]{revtex4}
\usepackage{graphicx}
\usepackage{dcolumn}
\usepackage{bm}

\begin{document}
\title{Circumstantial evidence for a non-Maxwellian plasma \\ from femtosecond laser-matter interaction}
\author{Sachie Kimura}
\author{Aldo Bonasera}
 \affiliation{INFN-LNS, via Santa Sofia, 62, 95123 Catania, Italy}
 \affiliation{Cyclotron Institute, Texas A\&M University, College Station TX 77843-3366, USA}

\date{\today}

\begin{abstract}
We study ion acceleration mechanisms in laser-plasma interactions using neutron spectroscopy.
We consider different types
of ion-collision mechanisms in the plasma, which cause the angular anisotropy of the observed
neutron spectra. These include the collisions between an ion in the plasma and an ion in the target, 
and the collisions between two ions in the hot plasma.
By analyzing the proton spectra,
we suggest that the laser-generated plasma 
consists of at least two components, one of which collectively accelerated and
can also produce anisotropy in the angular distribution of fusion neutrons.
\end{abstract}
\maketitle

\section{Introduction}

The development of small-scale high-intensity laser-systems with the chirped-pulse amplification~(CPA) 
technique has opened new research fields of the laser-plasma interaction~\cite{ledingham}.       
One of the promising applications of these studies is  
the ion-beam generation from laser irradiation on solid targets
\cite{PhysRevLett.88.215006,PhysRevE.67.046402,lee:056403}. 
It is reported that protons in Mylar~(H$_8$C$_{10}$O$_4$) target irradiation are accelerated more effectively
than in foil target irradiation.   
Understanding the ion acceleration mechanism in the laser-generated plasma is essential for applications.
In this connection, nuclear reactions induced by laser-irradiation give a unique clue in understanding 
the ion-acceleration mechanism~\cite{PhysRevLett.91.015001}.  
By replacing the protons in the plastic CH$_n$ target by deuterons (CD$_n$), 
a plasma of deuterium ions is generated. 
In the plasma the reaction D($d,n$)$^3$He with a $Q$-value of 3.26 MeV is induced~\cite{norreys,izumi,habara} 
and produces monochromatic neutrons.
The angular distribution of the neutrons shows peculiar anisotropy
not only on CD-plastic target~\cite{hilscher,izumi,habara} but also on D$_2$-gas jet~\cite{fritzler}
and on both D$_2$ and CD$_2$ clusters~\cite{ditmire,buersgens}.
The observed neutron angular distribution gives a direct hint to understand ion acceleration
mechanisms in aneutronic reactions driven by laser as well~\cite{belyaev:026406,kab,abc}.    

In a recent paper \cite{habara}, Habara et al. discuss the results of an experimental analysis of the 
neutron spectra in nuclear reactions induced by a laser-irradiation on a plastic CD target 50 $\mu$m thick.  
They observed that neutron counts at $I$=1 $\times$10$^{19}$ W/cm$^2$ is larger and more anisotropic 
than that at $I$=2 $\times$ 10$^{18}$ W/cm$^2$.
This is attributed to the fact that higher intensity laser-pulse can accelerate ions more efficiently.
The ion temperatures of 70 and 300 keV
at $I$=2 $\times$10$^{18}$ and 1$\times$ 10$^{19}$ W/cm$^2$, respectively, with a
similar number of accelerated ions ($N_i$=10$^{13}$) for both intensities are deduced,
using a three-dimensional Monte Carlo (3D MC) code. 
They simulate ion acceleration processes 
under different assumptions and conclude that 
the deuterium ions are accelerated into the target
and cause the nuclear reaction in the target.
The directionality of the plasma beam is deduced from the comparison to the differential cross section 
data of the reaction D($d,n$)$^3$He in conventional laboratory beam-target experiments~\cite{hu49,th66,schulte,li73,ja77,brown}.  
In this paper
we consider different types
of ion-collision mechanisms in plasma, which cause the angular anisotropy of the observed
neutron spectra, including collisions between an ion in the hot plasma and an ion in the target~(HT), 
and collisions between two ions in the hot plasma~(HH),
using at first the total number of accelerated ions and the plasma temperature 
given in Ref.~\cite{habara}.
This assumption results in the overestimate of the absolute value of the neutron yield, 
if fusion for collisions between ions in the hot plasma are properly included. This component was ignored in 
Ref.~\cite{habara}.
Using the SRIM code~\cite{srim} we estimate the HT component which suggests a smaller number of plasma ions
compared to Ref.~\cite{habara}.  This, in turn, reduces the number of fusion originating from the collisions among hot ions in the plasma reconciling to the experimental observation.  However, this is not the only possible explanation for the observed
angular anisotropy in the neutron data.  In fact the angular anisotropy in the neutron spectra can be observed, if a part of
the plasma is collectively accelerated and even in the absence of HT mechanism.  In order to shed some light on this point 
we study the plasma distribution reported in~\cite{PhysRevE.67.046402,lee:056403} and show that indeed such
a collective component is observed.
We mention that the origin and the mechanism of accelerated ions have been discussed in detail, for a review
Ref.~\cite{fuchs,PhysRevLett.91.015001}, 
and by now it is known the existence of at least two types of ion acceleration mechanisms, i.e., from the target front 
side into the target and from the target rear side to the vacuum. One of these mechanism becomes predominant depending on the 
target material and thickness or laser parameters. The later can be a candidate for the collectively accelerated plasma.   
Finally, we stress the importance of knowing both characteristics of fusion product, i.e., the spectra of neutron yield and the spectra of plasma ions, under common experimental conditions. 
At present those data are available only separately.

\section{Number of plasma ions derived from the observed neutron yield}
\label{sec:theo}

In practice we consider the following two types of mechanisms for neutron generation in high intensity laser irradiation. 
\begin{itemize}
\item[(A)]  Collisions between two ions in the laser-heated plasma. Both ions are moving with thermal velocity. Under this assumption 
the direction of the incident reaction channel is random, hence the angular 
distribution of reaction products will be isotropic~\cite{izumi}. 
The contribution to the neutron yield from this mechanism is called ``HH'' in this paper.
\item[(B)]  Collisions between an accelerated ion in the laser-produced plasma and a cold nucleus in the bulk of the target.
Under this second assumption the angular distribution of reaction products is possibly anisotropic. The contribution 
from this mechanism is called ``HT'' component. 
\end{itemize}
We stress the importance of considering both mechanisms mentioned above, comprehensively, because 
either mechanism might be predominant, depending on the characteristics of the target.   
As an example in the case of neutron yield observation from laser pulses irradiation on deuterated clusters~\cite{buersgens}
both mechanisms play a key role.  

If we assume that neutrons are produced by the collision of the ions in the hot plasma component (HH),
in terms of the number of the accelerated ions $N_i$,    
the number of fusion per solid angle, or reaction rate~\cite{clayton}, is given by 
\begin{equation}
  \label{eq:nfh}
  \frac{N_f^{(HH)}}{4\pi}=\frac{1}{4\pi}N_i n_{cr} \tau \int \sigma(v)v \phi(v)dv^3,
\end{equation}
where $n_{cr}=$10$^{21}$/$\lambda^2$ is the plasma critical density~\cite{forslund, forslund2}; 
$\tau$ is the laser pulse duration;
$\sigma(v)$ and $v$ are the reaction cross section and the relative velocity of the colliding ions. 
In general the reaction cross section $\sigma(v)$ is given as a function of the incident energy, instead of the velocity, but here we have written it as a function of the velocity to keep the consistency in the velocity integral. Later $\sigma$ will be represented as a function of the incident energy.  
We stress that assuming the critical density, which is the lowest limit to the real density reached 
in the experiment, the fusion yield given by Eq.(\ref{eq:nfh}) is underestimated. The density profile 
of the plasma simulated in Ref~\cite{habara}, using the PIC code, has an exponential shape which varies from  
4$n_{cr}$ to 0.1$n_{cr}$. 
$\phi(v)$ is the relative velocity spectrum of a pair of ions and is given by  
a Maxwellian-distribution at the temperature $kT_{HH}=$70 or 300~keV:
\begin{equation}
 \phi(v)=\left(\frac{\mu}{2\pi kT_{HH}} \right)^{\frac{3}{2}}\exp \left(-\frac{\mu v^2}{2kT_{HH}} \right), \label{eq:vs} 
\end{equation}
where $\mu$ is the reduced mass of ions. 
Eq.~(\ref{eq:nfh}) gives 1.$\times$ 10$^4$ and 4.$\times$10$^4$, per solid angle, at the 
temperature of 70 and of 300~keV, respectively, see Tab.\ref{tab:pr1}.
The estimated yield is comparable with the neutron spectra in Fig.s 3 and 5 in Ref. \cite{habara}. 
The contribution from HH component is, therefore, expected to be seen in the figure as a peak 
at the neutron energy 2.45 MeV and to be isotropic \cite{izumi}. 
In order to compare this to their energy distribution, we can roughly assume that the neutron distribution is a Gaussian distribution with a center-of-mass~(CM) energy at 2.45 MeV and a width given by the temperature of the plasma.  This gives an estimate at all angles of 1.4 $\times$10$^5$ and 1.3 $\times$ 10$^5$ ion/MeV/sr respectively, which is seen neither in figures 3 nor 5 in Ref. \cite{habara}.
In the figures, if anything, one sees a clear angular dependence of the neutron yield and 
shifts of the observed peaks from the expected energy 2.45 MeV.  
This implies that either their estimated temperature or the number of accelerated ions given is too large.  In other 
words the authors of Ref.~\cite{habara} should have estimated the number of neutrons coming from the HH component, 
and show that this component is negligible compared to the HT contribution.  
As we will show in the following discussion, 
the contribution from the HT component, which has the correct angular dependence of the observed neutron spectra, is indeed dominant 
but with a smaller number of plasma ions. 

If we consider 
the collisions between the ions in the plasma and the almost stable nuclei in the target (HT),
the angular distribution of reaction products is expected to be anisotropic.
To estimate the neutron yield from the HT component, one should take into account that 
one of the colliding ions is at rest in the laboratory frame.
Therefore in the reaction rate per pair of colliding ions, 
the velocity spectrum Eq.(\ref{eq:vs}) is modified as: 
\begin{equation}
 \phi_{HT}(v)=\left(\frac{m_1}{2\pi kT_{HH}} \right)^{\frac{3}{2}}\exp \left(-\frac{m_1 v^2}{2kT_{HH}} \right), \label{eq:vsHT} 
\end{equation}
where $m_1$ is the mass of ions in plasma, instead of the reduced mass. 
One can define an
effective temperature as, 
\begin{equation}
 kT^{eff}_{HT}=(\mu/m_1)kT_{HH}.  \label{eq:effectt}
\end{equation}
Now we can use the effective temperature to estimate the number of fusion.
For simplicity we estimate the most probable energy of the plasma ions that cause the nuclear reaction 
given by the Gamow peak energy 
($E_G$) \cite{clayton}. The Gamow energy
can be found using
the saddle point method, i.e.:
\begin{equation}
  \label{eq:gamow}
  \frac{d}{dE}\left(\frac{E}{kT}+bE^{-\frac{1}{2}} \right)=0,
\end{equation}
where $b=31.28Z_1Z_2 M^{\frac{1}{2}}$ (keV$^{\frac{1}{2}}$), denoting the atomic number of the colliding nuclei $Z_1, Z_2$ and 
the reduced mass number $M$. We remind that $M=A_1A_2/(A_1+A_2)$, where $A_1$ and $A_2$ are the mass numbers of the colliding 
ions, respectively, is different from the reduced mass $\mu$. 
The temperature $kT$ is replaced by the plasma temperature, i.e., 
the temperature of the HH component in conventional discussions, 
but for the present case of the HT component, 
$kT$ is replaced by the effective temperature.
Then, 
\begin{equation}
  \label{eq:gamow2}
  E_G=\left(bkT^{eff}_{HT}/2\right)^{\frac{2}{3}}. 
\end{equation}
We specially mention that Eq.~(\ref{eq:gamow}) is valid in the case of sub-barrier 
reactions, because this condition comes from the product of a Maxwellian and the Coulomb barrier 
penetrability~\cite{clayton}. The height of the Coulomb barrier for the reaction D($d,n$)$^3$He is estimated to be about 470 keV.  
By approximating the energy of the accelerated ions by the Gamow energy Eq.~(\ref{eq:gamow2}), 
the neutron yield per solid angle is 
written as:  
\begin{equation}
  \label{eq:nat2}
  \frac{N_f^{(HT)}}{4\pi} =\frac{1}{4\pi} N_i \sigma(E_G) n_T d,       
\end{equation}
where $\sigma(E_G)$ is the reaction cross section at the Gamow energy.
The reaction cross section data at this energy are taken from NACRE compilation~\cite{nacre} and 
given in Table~\ref{tab:pr1}.  
The number density of the solid polyethylene target: $n_T=$ 1.2 $\times$ 10$^{23}$ atoms/cm$^3$ and $d$ is the projected range \cite{srim} of the accelerated ions in the target.
Using the above listed numbers the neutron yield per solid angle is also reported in Tab.~\ref{tab:pr1}.     
\begin{table*}
\caption{Plasma temperature ($kT_{HH}$) and neutron yield per solid angle ($N_f^{(HH)}/Sr$) from the HH component and 
the effective temperature ($kT^{eff}_{HT}$) and $N_f^{(HT)}/Sr$ from the HT component at the given laser intensities.}
\label{tab:pr1}
\begin{ruledtabular}
\begin{tabular}{c|cc|ccccc}
I (W/cm$^2$)&  $kT_{HH}$ (keV) & $N_f^{(HH)}/Sr$ & $kT^{eff}_{HT}$(keV) & $E_G$(keV) & $\sigma(E_G)$(10$^{-27}$cm$^{2}$) & $d(\mu$m) & $N_f^{(HT)}/Sr$ \\
\hline 
2$\times$10$^{18}$ & 70         & 1.$\times$10$^4$ & 35       &  67  &  24    &  0.8   & 4.$\times$10$^5$    \\
1$\times$10$^{19}$ & 300        & 4.$\times$10$^4$ & 150      &  176  &  66    &  2.   & 3.$\times$10$^6$       
\end{tabular}
\end{ruledtabular}
\end{table*}
As a consequence of this simple estimate, the yield per solid angle under this assumption is a factor 10$^2$ higher than 
the yield from the collisions between two ions in the plasma.
This is because the number density of the solid target is much higher than that of the laser-generated plasma.  
However from eq.~(\ref{eq:effectt}) the effective HT temperature is lower than the HH one, depending on the asymmetry in mass of the fusion ions. The higher temperature in the HH collision mechanism has an advantage of the higher fusion cross section. 
There is, therefore, a competition between two mechanism depending on the laser-intensity.
This feature might explain why at lower intensities (i.e. lower T) 
the neutron angular distribution is less anisotropic:
At the lower intensity irradiation,  the contribution from the HT component to the neutron yield 
is suppressed compared with the higher intensity irradiation. 
In fact the temperature of the HT component is low and most reactions happen below the Coulomb barrier.
The nuclear reactions below the barrier are exponentially suppressed.  Therefore the HH mechanism, which has higher temperature, 
contributes to the less anisotropic neutron yield angular distribution. 
On the other hand, if the plasma temperature is high enough, the corresponding Gamow energy is above the Coulomb barrier. 
 At energies above the Coulomb barrier, i.e. high T, to have higher densities gives higher fusion probabilities (above the barrier fusion probabilities depend quadratically on densities and do not depend much on T) thus collisions between an ion from the plasma and an ion from the target becomes more probable.  
 We expect that in some temperature region there is a transition from collisions occurring mainly in the plasma to collisions occurring between the plasma ions and the target ones. An experimental detailed investigation of this transition region would be very interesting and instructive in order to understand the microscopic dynamics of fusion.

The HT contribution to the neutron yield is a factor 10$^2$ higher than experimentally observed data~\cite{habara}, as well.  
This again implies that either the estimated temperatures are too high or the estimated number 
of ions in the plasma is too large. 
From this simple estimate and the one above from the HH component we can argue 
that the number of ions in the plasma could be at least two orders of magnitude smaller than that estimated in Ref.~\cite{habara}, i.e., about 10$^{11}$, instead of 10$^{13}$. This reduction of the number of accelerated ions results in a suppression of the neutron yield from the HH component
which becomes negligible. 
A more involved calculation solving eq.(1) for the HT component numerically gives results in agreement with our simple estimate and will be discussed in more detail in a following paper~\cite{kb-t}.

\section{Plasma temperature suggested by the proton spectra from plastic targets irradiation}
\label{sec:psp}

Another evidence of the smaller number of plasma ions is shown by a direct observation of proton 
spectra in laser irradiation~\cite{PhysRevE.67.046402,lee:056403}.
Fig. 1 in Lee et al.~\cite{lee:056403} shows the proton energy distribution
at the laser intensity $I=$ 2.2$\times$ 10$^{18}$ W/cm$^2$ similar to~\cite{habara},
and the target material is Mylar or aluminum. 
We are especially interested in the results from the Mylar target,
because the characteristics of the produced plasma should be close to 
the characteristics of the plasma from the deuterated plastic targets. We, therefore, 
selected two results from 6 and 13~$\mu$m thick Mylar target.  
\begin{figure}
  \includegraphics[height=.24\textheight]{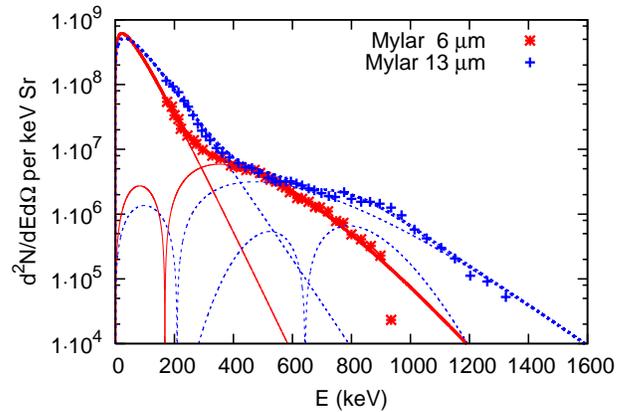}

  \caption{\label{fig:pes} Experimental data of proton energy spectra for Mylar targets 6~(asterisks) and 13~(crosses) $\mu$m-thick
    retrieved from Lee et al.~\cite{lee:056403}. The thick curves represent fitting of the proton spectra using Maxwellian distributions and distributions with extra accelerations. The thin curves are the different components and are summed into the thick curves. The obtained fitting parameters are given in Tab.~\ref{tab:fit}.}
\end{figure}
A peculiar feature of the proton energy distribution is that it exhibits bumps in the higher energy region. 
This feature could be attributed to the existence of at least two different components in the plasma.
To describe these characteristics, we use a Maxwellian distribution in terms of the energy of the protons 
in the lab. system ($E$): 
\begin{equation}
  \label{eq:nn}
  \frac{d^2 N}{dEd\Omega}=c\frac{1}{2\sqrt{\pi^3}}\frac{\sqrt{E}}{(kT_{HH})^{3/2}} \exp(-E/kT_{HH}).
\end{equation}
To this we add a Maxwellian distribution of a moving source~\cite{PhysRevC.24.89}:
\begin{equation}
  \label{eq:plf}
  \frac{d^2 N}{dEd\Omega}=\frac{c_1}{4\pi}\frac{\sqrt{EE'}}{(kT_1)^{2}} \exp(-E'/kT_1), 
\end{equation}
where $E'=E-2\sqrt{E E_0}+E_0$.
Both equations are normalized to 1/4$\pi$.
The latter gives a part of the plasma in translational motion with respect to the laboratory system with a collective 
kinetic energy $E_0$. 
The obtained fitting curves are shown by thick solid and dashed 
curves for the spectra at 6 and 13 $\mu$m thick targets irradiation, respectively, in Fig.~\ref{fig:pes}.
The thin curves show each component which add into the thick curves.
Especially for the spectra at 13 $\mu$m thick target, we have used a sum of three components, 
two of them have different extra acceleration energies $E_0$.  
The corresponding fitting parameters 
are summarized in Tab.~\ref{tab:fit}. 
\begin{table}
  \caption{Fitting parameters for proton energy spectra from 6 and 13 $\mu$m thick targets irradiation.
  Especially for 13 $\mu$m thick target we have used three components, two of which have different relative velocities with respect to 
  the hot-plasma component.}
  \label{tab:fit}
\begin{ruledtabular}
  \begin{tabular}{c|ccccc}
    T thick. & $c$ & $kT_{HH}$  & $c_1$ & $kT_1$ & $E_0$  \\
    ($\mu$m) & (10$^{11}$) & (keV) &  & (keV) & (keV)   \\
    \hline
    6 & (3.0$\pm$0.2)  & 44.$\pm$1.  & (3.5$\pm$0.7) 10$^{9}$ & 52.$\pm$5. & 170.$\pm$20.  \\
    13 & (3.5$\pm$0.1)  & 61.$\pm$1. & (3. $\pm$ 2.) 10$^{9}$ & 78.$\pm$30. & 210.$\pm$60 \\
       &                           &            & (4. $\pm$ 1.) 10$^{7}$ & 14.$\pm$5.  & 695.$\pm$10  \\ 
  \end{tabular}
\end{ruledtabular}
\end{table}
From the fitting we can deduce the number of ions in the plasma $N_i\sim$ 3$\times$10$^{11}$ 
and the plasma temperature of $kT_{HH}\sim$ 44 and 61 keV for 6 and 13 $\mu$m thick targets, respectively.  
The deduced number of plasma ions is in agreement with 
our yield estimation from the neutron measurement.    
Moreover one can see clearly that the proton spectra differ from a Maxwellian 
but a part of plasma is accelerated to higher energy with $E_0$.
Similar features of the proton spectra in laser-produced plasma can be found in Spencer et al.~\cite{PhysRevE.67.046402}, as well.
The possible reasons for such a collective acceleration will be discussed in a following publication~\cite{kb-t}.

\begin{figure}
  \includegraphics[height=.24\textheight]{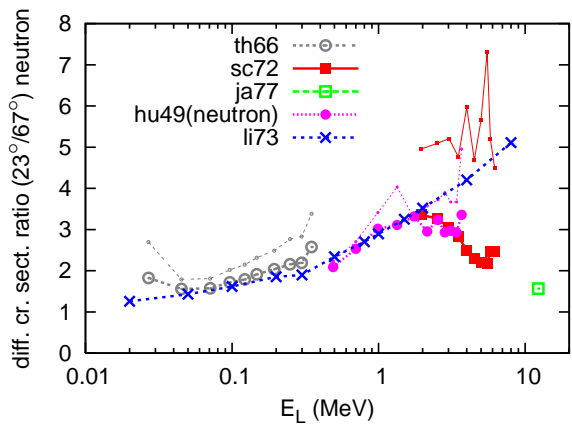}

  \caption{\label{fig:Dd_anglratioN} The ratio of differential cross sections at two angles.  
The ratio at angles 23$^{\circ}$ to 67$^{\circ}$ is shown by large points connected by curves, while the ratio at angles 20$^{\circ}$ to 85$^{\circ}$ is shown by smaller points connected by thin curves, for a comparison. 
The experimental data are retrieved from EXFOR-data system \cite{exfor}. hu49-data~(pink filled circles) are evaluated from the angular distribution of the DD neutron yield~\cite{hu49} and the others, th66-data~(grey open circles)~\cite{th66}, sc72~(red filled squares)~\cite{schulte}, ja77~(green open squares)~\cite{ja77} and li73~(blue crosses)~\cite{li73}, are converted from the $^3$He angular distribution of the DD reaction. }
\end{figure}

In Fig. 7 and 8 of~\cite{habara}, the authors discuss the ratio of the neutron yields at the 
angle of 23$^{\circ}$ to that at the angle of 67$^{\circ}$ from the target rear normal. 
They found that the ratio can be larger than 3.
%
%
In Fig.~\ref{fig:Dd_anglratioN} the ratio of the differential cross sections in beam-target experiments\cite{hu49,th66,schulte,li73,ja77,brown}
at laboratory angles 23$^{\circ}$ to 67$^{\circ}$ is shown by thick curves.
There is a discrepancy between [sc72] data~\cite{schulte} and [li73]~\cite{li73} in the energy region higher than 
3 MeV. Nevertheless the figure shows clearly that the ratio can, indeed, be larger than 3 and
less than 3.5 at the incident deuteron energy $E_L$ from 1.75 to 1.96~MeV. 
%
Thus the accelerated deuterons which contribute 
to the nuclear reactions have energies slightly less than 2~MeV. 
In the CM frame $E\sim$0.9~MeV, i.e., the corresponding Gamow peak energy is about 900~keV.
If we use $E_G$=900~keV, the neutron yield 
per solid angle from the HT collision is estimated to be of the order of 10$^7$ from Eq.~(\ref{eq:nat2}), 
adopting the number of accelerated ions given in~\cite{habara}. This number is 10 times higher than that 
in Table~\ref{tab:pr1}. In other words, to reproduce the experimental data, one needs 
to assume $N_i\sim$10$^{10}$, as the number of accelerated ions in plasma.
In passing we note that the ratio of the neutron yield taken at the angle of 20$^{\circ}$ to that of 85$^{\circ}$ is higher 
than the previous one as shown in Fig.~\ref{fig:Dd_anglratioN} by thin curves with smaller points.
A careful experimental determination of this feature would be very useful. 


We can also derive the plasma temperature corresponding to $E_G$=900~keV. 
Since at this energy we are above the Coulomb barrier, we can estimate it from the classical relation:

\begin{equation}
  \label{eq:egtemp}
  E_G=(3/2)kT_{HH},
\end{equation}
which gives $kT_{HH}$=600~keV, i.e., 
higher than the estimated temperature in Ref. \cite{habara}. 
This contradiction might be solved 
by considering that a part of the plasma deuterons is collectively accelerated at energy $E_0$, 
as we have shown in the analysis of proton spectra. 
This extra acceleration energy can reach about 700~keV
which is close to 900~keV. 
In the presence of an extra acceleration,   
the plasma temperature cannot be derived simply from the relation (\ref{eq:egtemp}). 
The difference between the plasma temperature
deduced from the ratio at two angles and the one estimated by Ref.~\cite{habara} 
suggests that the plasma spectra is different from a usual Maxwellian distribution. 
In other words, the difference justifies 
the presence of the collective motion of a part of the plasma.      
Indeed the collisions between two ions in the plasma which is moving at a collective energy $E_0$ 
in the laboratory frame can result in the angular anisotropy of the neutron yield.   
This is a possible mechanism to explain the angular anisotropy experimentally observed.

\section{conclusions}
\label{sec:conc}

In conclusion,
we have discussed different possible ion-collision mechanisms, 
which can result in the observed anisotropic neutron spectra.
When analyzing the proton spectra, we have found that there are at least two plasma components: one is 
approximated by a Maxwellian distribution and the other has a collective motion relative to the former component.  
Comparing the ratios in the neutron counts at two angles in the experiment
to the differential cross sections measured in conventional beam-target experiments,
the most effective energy is estimated to be 0.9 MeV
with corresponding plasma temperature of 600 keV, 
which is higher than the estimated plasma temperature in Ref.~\cite{habara}. 
We have discussed a possible solution of this contradiction, in connection with the collective
motion of a part of the plasma, which can in principle explain the observed neutron angular
anisotropy.
We suggest that an experimental determination of the neutron angular anisotropy in coincidence with the plasma distribution would be very interesting and give useful information on the mechanisms at play.

\bibliography{memo2}

\begin{thebibliography}{30}
\expandafter\ifx\csname natexlab\endcsname\relax\def\natexlab#1{#1}\fi
\expandafter\ifx\csname bibnamefont\endcsname\relax
  \def\bibnamefont#1{#1}\fi
\expandafter\ifx\csname bibfnamefont\endcsname\relax
  \def\bibfnamefont#1{#1}\fi
\expandafter\ifx\csname citenamefont\endcsname\relax
  \def\citenamefont#1{#1}\fi
\expandafter\ifx\csname url\endcsname\relax
  \def\url#1{\texttt{#1}}\fi
\expandafter\ifx\csname urlprefix\endcsname\relax\def\urlprefix{URL }\fi
\providecommand{\bibinfo}[2]{#2}
\providecommand{\eprint}[2][]{\url{#2}}

\bibitem[{\citenamefont{Ledingham et~al.}(2003)\citenamefont{Ledingham,
  McKenna, and Singhal}}]{ledingham}
\bibinfo{author}{\bibfnamefont{K.}~\bibnamefont{Ledingham}},
  \bibinfo{author}{\bibfnamefont{P.}~\bibnamefont{McKenna}}, \bibnamefont{and}
  \bibinfo{author}{\bibfnamefont{R.}~\bibnamefont{Singhal}},
  \bibinfo{journal}{Science} \textbf{\bibinfo{volume}{300}},
  \bibinfo{pages}{1107} (\bibinfo{year}{2003}).

\bibitem[{\citenamefont{Mackinnon et~al.}(2002)\citenamefont{Mackinnon,
  Sentoku, Patel, Price, Hatchett, Key, Andersen, Snavely, and
  Freeman}}]{PhysRevLett.88.215006}
\bibinfo{author}{\bibfnamefont{A.~J.} \bibnamefont{Mackinnon}},
  \bibinfo{author}{\bibfnamefont{Y.}~\bibnamefont{Sentoku}},
  \bibinfo{author}{\bibfnamefont{P.~K.} \bibnamefont{Patel}},
  \bibinfo{author}{\bibfnamefont{D.~W.} \bibnamefont{Price}},
  \bibinfo{author}{\bibfnamefont{S.}~\bibnamefont{Hatchett}},
  \bibinfo{author}{\bibfnamefont{M.~H.} \bibnamefont{Key}},
  \bibinfo{author}{\bibfnamefont{C.}~\bibnamefont{Andersen}},
  \bibinfo{author}{\bibfnamefont{R.}~\bibnamefont{Snavely}}, \bibnamefont{and}
  \bibinfo{author}{\bibfnamefont{R.~R.} \bibnamefont{Freeman}},
  \bibinfo{journal}{Phys. Rev. Lett.} \textbf{\bibinfo{volume}{88}},
  \bibinfo{pages}{215006} (\bibinfo{year}{2002}).

\bibitem[{\citenamefont{Spencer et~al.}(2003)\citenamefont{Spencer, Ledingham,
  McKenna, McCanny, Singhal, Foster, Neely, Langley, Divall, Hooker
  et~al.}}]{PhysRevE.67.046402}
\bibinfo{author}{\bibfnamefont{I.}~\bibnamefont{Spencer}},
  \bibinfo{author}{\bibfnamefont{K.~W.~D.} \bibnamefont{Ledingham}},
  \bibinfo{author}{\bibfnamefont{P.}~\bibnamefont{McKenna}},
  \bibinfo{author}{\bibfnamefont{T.}~\bibnamefont{McCanny}},
  \bibinfo{author}{\bibfnamefont{R.~P.} \bibnamefont{Singhal}},
  \bibinfo{author}{\bibfnamefont{P.~S.} \bibnamefont{Foster}},
  \bibinfo{author}{\bibfnamefont{D.}~\bibnamefont{Neely}},
  \bibinfo{author}{\bibfnamefont{A.~J.} \bibnamefont{Langley}},
  \bibinfo{author}{\bibfnamefont{E.~J.} \bibnamefont{Divall}},
  \bibinfo{author}{\bibfnamefont{C.~J.} \bibnamefont{Hooker}},
  \bibnamefont{et~al.}, \bibinfo{journal}{Phys. Rev. E}
  \textbf{\bibinfo{volume}{67}}, \bibinfo{pages}{046402}
  (\bibinfo{year}{2003}).

\bibitem[{\citenamefont{Lee et~al.}(2008)\citenamefont{Lee, Park, Cha, Lee,
  Lee, Yea, and Jeong}}]{lee:056403}
\bibinfo{author}{\bibfnamefont{K.}~\bibnamefont{Lee}},
  \bibinfo{author}{\bibfnamefont{S.~H.} \bibnamefont{Park}},
  \bibinfo{author}{\bibfnamefont{Y.-H.} \bibnamefont{Cha}},
  \bibinfo{author}{\bibfnamefont{J.~Y.} \bibnamefont{Lee}},
  \bibinfo{author}{\bibfnamefont{Y.~W.} \bibnamefont{Lee}},
  \bibinfo{author}{\bibfnamefont{K.-H.} \bibnamefont{Yea}}, \bibnamefont{and}
  \bibinfo{author}{\bibfnamefont{Y.~U.} \bibnamefont{Jeong}},
  \bibinfo{journal}{Phys. Rev. E} \textbf{\bibinfo{volume}{78}},
  \bibinfo{eid}{056403} (\bibinfo{year}{2008}).

\bibitem[{\citenamefont{Karsch et~al.}(2003)\citenamefont{Karsch, D\"usterer,
  Schwoerer, Ewald, Habs, Hegelich, Pretzler, Pukhov, Witte, and
  Sauerbrey}}]{PhysRevLett.91.015001}
\bibinfo{author}{\bibfnamefont{S.}~\bibnamefont{Karsch}},
  \bibinfo{author}{\bibfnamefont{S.}~\bibnamefont{D\"usterer}},
  \bibinfo{author}{\bibfnamefont{H.}~\bibnamefont{Schwoerer}},
  \bibinfo{author}{\bibfnamefont{F.}~\bibnamefont{Ewald}},
  \bibinfo{author}{\bibfnamefont{D.}~\bibnamefont{Habs}},
  \bibinfo{author}{\bibfnamefont{M.}~\bibnamefont{Hegelich}},
  \bibinfo{author}{\bibfnamefont{G.}~\bibnamefont{Pretzler}},
  \bibinfo{author}{\bibfnamefont{A.}~\bibnamefont{Pukhov}},
  \bibinfo{author}{\bibfnamefont{K.}~\bibnamefont{Witte}}, \bibnamefont{and}
  \bibinfo{author}{\bibfnamefont{R.}~\bibnamefont{Sauerbrey}},
  \bibinfo{journal}{Phys. Rev. Lett.} \textbf{\bibinfo{volume}{91}},
  \bibinfo{pages}{015001} (\bibinfo{year}{2003}).

\bibitem[{\citenamefont{Norreys et~al.}(1998)\citenamefont{Norreys, Fews, Beg,
  Bell, Dangor, Lee, Nelson, Schmidt, Tatarakis, and Cable}}]{norreys}
\bibinfo{author}{\bibfnamefont{P.~A.} \bibnamefont{Norreys}},
  \bibinfo{author}{\bibfnamefont{A.~P.} \bibnamefont{Fews}},
  \bibinfo{author}{\bibfnamefont{F.~N.} \bibnamefont{Beg}},
  \bibinfo{author}{\bibfnamefont{A.~R.} \bibnamefont{Bell}},
  \bibinfo{author}{\bibfnamefont{A.~E.} \bibnamefont{Dangor}},
  \bibinfo{author}{\bibfnamefont{P.}~\bibnamefont{Lee}},
  \bibinfo{author}{\bibfnamefont{M.~B.} \bibnamefont{Nelson}},
  \bibinfo{author}{\bibfnamefont{H.}~\bibnamefont{Schmidt}},
  \bibinfo{author}{\bibfnamefont{M.}~\bibnamefont{Tatarakis}},
  \bibnamefont{and} \bibinfo{author}{\bibfnamefont{M.~D.} \bibnamefont{Cable}},
  \bibinfo{journal}{Plasma Phys. Control. Fusion}
  \textbf{\bibinfo{volume}{40}}, \bibinfo{pages}{175} (\bibinfo{year}{1998}).

\bibitem[{\citenamefont{Izumi et~al.}(2002)\citenamefont{Izumi, Sentoku,
  Habara, Takahashi, Ohtani, Sonomoto, Kodama, Norimatsu, Fujita, Kitagawa
  et~al.}}]{izumi}
\bibinfo{author}{\bibfnamefont{N.}~\bibnamefont{Izumi}},
  \bibinfo{author}{\bibfnamefont{Y.}~\bibnamefont{Sentoku}},
  \bibinfo{author}{\bibfnamefont{H.}~\bibnamefont{Habara}},
  \bibinfo{author}{\bibfnamefont{K.}~\bibnamefont{Takahashi}},
  \bibinfo{author}{\bibfnamefont{F.}~\bibnamefont{Ohtani}},
  \bibinfo{author}{\bibfnamefont{T.}~\bibnamefont{Sonomoto}},
  \bibinfo{author}{\bibfnamefont{R.}~\bibnamefont{Kodama}},
  \bibinfo{author}{\bibfnamefont{T.}~\bibnamefont{Norimatsu}},
  \bibinfo{author}{\bibfnamefont{H.}~\bibnamefont{Fujita}},
  \bibinfo{author}{\bibfnamefont{Y.}~\bibnamefont{Kitagawa}},
  \bibnamefont{et~al.}, \bibinfo{journal}{Phys. Rev. E}
  \textbf{\bibinfo{volume}{65}}, \bibinfo{pages}{036413}
  (\bibinfo{year}{2002}).

\bibitem[{\citenamefont{Habara et~al.}(2004)\citenamefont{Habara, Kodama,
  Sentoku, Izumi, Kitagawa, Tanaka, Mima, and Yamanaka}}]{habara}
\bibinfo{author}{\bibfnamefont{H.}~\bibnamefont{Habara}},
  \bibinfo{author}{\bibfnamefont{R.}~\bibnamefont{Kodama}},
  \bibinfo{author}{\bibfnamefont{Y.}~\bibnamefont{Sentoku}},
  \bibinfo{author}{\bibfnamefont{N.}~\bibnamefont{Izumi}},
  \bibinfo{author}{\bibfnamefont{Y.}~\bibnamefont{Kitagawa}},
  \bibinfo{author}{\bibfnamefont{K.~A.} \bibnamefont{Tanaka}},
  \bibinfo{author}{\bibfnamefont{K.}~\bibnamefont{Mima}}, \bibnamefont{and}
  \bibinfo{author}{\bibfnamefont{T.}~\bibnamefont{Yamanaka}},
  \bibinfo{journal}{Phys. Rev. E} \textbf{\bibinfo{volume}{69}},
  \bibinfo{pages}{036407} (\bibinfo{year}{2004}).

\bibitem[{\citenamefont{Hilscher et~al.}(2001)\citenamefont{Hilscher, Berndt,
  Enke, Jahnke, Nickles, Ruhl, and Sandner}}]{hilscher}
\bibinfo{author}{\bibfnamefont{D.}~\bibnamefont{Hilscher}},
  \bibinfo{author}{\bibfnamefont{O.}~\bibnamefont{Berndt}},
  \bibinfo{author}{\bibfnamefont{M.}~\bibnamefont{Enke}},
  \bibinfo{author}{\bibfnamefont{U.}~\bibnamefont{Jahnke}},
  \bibinfo{author}{\bibfnamefont{P.~V.} \bibnamefont{Nickles}},
  \bibinfo{author}{\bibfnamefont{H.}~\bibnamefont{Ruhl}}, \bibnamefont{and}
  \bibinfo{author}{\bibfnamefont{W.}~\bibnamefont{Sandner}},
  \bibinfo{journal}{Phys. Rev. E} \textbf{\bibinfo{volume}{64}},
  \bibinfo{pages}{016414} (\bibinfo{year}{2001}).

\bibitem[{\citenamefont{Fritzler et~al.}(2002)\citenamefont{Fritzler, Najmudin,
  Malka, Krushelnick, Marle, Walton, Wei, Clarke, and Dangor}}]{fritzler}
\bibinfo{author}{\bibfnamefont{S.}~\bibnamefont{Fritzler}},
  \bibinfo{author}{\bibfnamefont{Z.}~\bibnamefont{Najmudin}},
  \bibinfo{author}{\bibfnamefont{V.}~\bibnamefont{Malka}},
  \bibinfo{author}{\bibfnamefont{K.}~\bibnamefont{Krushelnick}},
  \bibinfo{author}{\bibfnamefont{C.}~\bibnamefont{Marle}},
  \bibinfo{author}{\bibfnamefont{B.}~\bibnamefont{Walton}},
  \bibinfo{author}{\bibfnamefont{M.~S.} \bibnamefont{Wei}},
  \bibinfo{author}{\bibfnamefont{R.~J.} \bibnamefont{Clarke}},
  \bibnamefont{and} \bibinfo{author}{\bibfnamefont{A.~E.}
  \bibnamefont{Dangor}}, \bibinfo{journal}{Phys. Rev. Lett.}
  \textbf{\bibinfo{volume}{89}}, \bibinfo{pages}{165004}
  (\bibinfo{year}{2002}).

\bibitem[{\citenamefont{Ditmire et~al.}(1999)\citenamefont{Ditmire, Zweiback,
  Yanovsky, Cowan, Hays, and Wharton}}]{ditmire}
\bibinfo{author}{\bibfnamefont{T.}~\bibnamefont{Ditmire}},
  \bibinfo{author}{\bibfnamefont{J.}~\bibnamefont{Zweiback}},
  \bibinfo{author}{\bibfnamefont{V.}~\bibnamefont{Yanovsky}},
  \bibinfo{author}{\bibfnamefont{T.}~\bibnamefont{Cowan}},
  \bibinfo{author}{\bibfnamefont{G.}~\bibnamefont{Hays}}, \bibnamefont{and}
  \bibinfo{author}{\bibfnamefont{K.}~\bibnamefont{Wharton}},
  \bibinfo{journal}{Nature} \textbf{\bibinfo{volume}{398}},
  \bibinfo{pages}{489} (\bibinfo{year}{1999}).

\bibitem[{\citenamefont{Buersgens et~al.}(2006)\citenamefont{Buersgens,
  Madison, Symes, Hartke, Osterhoff, Grigsby, Dyer, and Ditmire}}]{buersgens}
\bibinfo{author}{\bibfnamefont{F.}~\bibnamefont{Buersgens}},
  \bibinfo{author}{\bibfnamefont{K.~W.} \bibnamefont{Madison}},
  \bibinfo{author}{\bibfnamefont{D.~R.} \bibnamefont{Symes}},
  \bibinfo{author}{\bibfnamefont{R.}~\bibnamefont{Hartke}},
  \bibinfo{author}{\bibfnamefont{J.}~\bibnamefont{Osterhoff}},
  \bibinfo{author}{\bibfnamefont{W.}~\bibnamefont{Grigsby}},
  \bibinfo{author}{\bibfnamefont{G.}~\bibnamefont{Dyer}}, \bibnamefont{and}
  \bibinfo{author}{\bibfnamefont{T.}~\bibnamefont{Ditmire}},
  \bibinfo{journal}{Phys. Rev. E} \textbf{\bibinfo{volume}{74}},
  \bibinfo{pages}{016403} (\bibinfo{year}{2006}).

\bibitem[{\citenamefont{Belyaev et~al.}(2005)\citenamefont{Belyaev, Matafonov,
  Vinogradov, Krainov, Lisitsa, Roussetski, Ignatyev, and
  Andrianov}}]{belyaev:026406}
\bibinfo{author}{\bibfnamefont{V.~S.} \bibnamefont{Belyaev}},
  \bibinfo{author}{\bibfnamefont{A.~P.} \bibnamefont{Matafonov}},
  \bibinfo{author}{\bibfnamefont{V.~I.} \bibnamefont{Vinogradov}},
  \bibinfo{author}{\bibfnamefont{V.~P.} \bibnamefont{Krainov}},
  \bibinfo{author}{\bibfnamefont{V.~S.} \bibnamefont{Lisitsa}},
  \bibinfo{author}{\bibfnamefont{A.~S.} \bibnamefont{Roussetski}},
  \bibinfo{author}{\bibfnamefont{G.~N.} \bibnamefont{Ignatyev}},
  \bibnamefont{and} \bibinfo{author}{\bibfnamefont{V.~P.}
  \bibnamefont{Andrianov}}, \bibinfo{journal}{Phys. Rev. E}
  \textbf{\bibinfo{volume}{72}}, \bibinfo{eid}{026406} (\bibinfo{year}{2005}).

\bibitem[{\citenamefont{Kimura et~al.}(2009)\citenamefont{Kimura, Anzalone, and
  Bonasera}}]{kab}
\bibinfo{author}{\bibfnamefont{S.}~\bibnamefont{Kimura}},
  \bibinfo{author}{\bibfnamefont{A.}~\bibnamefont{Anzalone}}, \bibnamefont{and}
  \bibinfo{author}{\bibfnamefont{A.}~\bibnamefont{Bonasera}},
  \bibinfo{journal}{Phys. Rev. E} \textbf{\bibinfo{volume}{79}},
  \bibinfo{pages}{038401} (\bibinfo{year}{2009}).

\bibitem[{\citenamefont{Bonasera et~al.}(2008)\citenamefont{Bonasera, Caruso,
  Strangio, Aglione, Anzalone, Kimura, Leanza, Spitaleri, Imme, Morelli
  et~al.}}]{abc}
\bibinfo{author}{\bibfnamefont{A.}~\bibnamefont{Bonasera}},
  \bibinfo{author}{\bibfnamefont{A.}~\bibnamefont{Caruso}},
  \bibinfo{author}{\bibfnamefont{C.}~\bibnamefont{Strangio}},
  \bibinfo{author}{\bibfnamefont{M.}~\bibnamefont{Aglione}},
  \bibinfo{author}{\bibfnamefont{A.}~\bibnamefont{Anzalone}},
  \bibinfo{author}{\bibfnamefont{S.}~\bibnamefont{Kimura}},
  \bibinfo{author}{\bibfnamefont{D.}~\bibnamefont{Leanza}},
  \bibinfo{author}{\bibfnamefont{A.}~\bibnamefont{Spitaleri}},
  \bibinfo{author}{\bibfnamefont{G.}~\bibnamefont{Imme}},
  \bibinfo{author}{\bibfnamefont{D.}~\bibnamefont{Morelli}},
  \bibnamefont{et~al.}, \bibinfo{journal}{Fission and properties of
  neutron-rich nuclei - Proceedings of the Fourth International Conference} p.
  \bibinfo{pages}{503} (\bibinfo{year}{2008}).

\bibitem[{\citenamefont{Hunter and Richards}(1949)}]{hu49}
\bibinfo{author}{\bibfnamefont{G.}~\bibnamefont{Hunter}} \bibnamefont{and}
  \bibinfo{author}{\bibfnamefont{H.}~\bibnamefont{Richards}},
  \bibinfo{journal}{Phys. Rev.} \textbf{\bibinfo{volume}{76}},
  \bibinfo{pages}{1445} (\bibinfo{year}{1949}).

\bibitem[{\citenamefont{R.B.Theus et~al.}(1966)\citenamefont{R.B.Theus,
  W.I.McGarry, and L.A.Beach}}]{th66}
\bibinfo{author}{\bibnamefont{R.B.Theus}},
  \bibinfo{author}{\bibnamefont{W.I.McGarry}}, \bibnamefont{and}
  \bibinfo{author}{\bibnamefont{L.A.Beach}}, \bibinfo{journal}{Nucl. Phys.}
  \textbf{\bibinfo{volume}{80}}, \bibinfo{pages}{273} (\bibinfo{year}{1966}).

\bibitem[{\citenamefont{Schulte et~al.}(1972)\citenamefont{Schulte, Cosack,
  Obst, and Weil}}]{schulte}
\bibinfo{author}{\bibfnamefont{R.}~\bibnamefont{Schulte}},
  \bibinfo{author}{\bibfnamefont{M.}~\bibnamefont{Cosack}},
  \bibinfo{author}{\bibfnamefont{A.}~\bibnamefont{Obst}}, \bibnamefont{and}
  \bibinfo{author}{\bibfnamefont{J.}~\bibnamefont{Weil}},
  \bibinfo{journal}{Nucl. Phys. A} \textbf{\bibinfo{volume}{192}},
  \bibinfo{pages}{609} (\bibinfo{year}{1972}).

\bibitem[{\citenamefont{Liskien and A.Paulsen}(1973)}]{li73}
\bibinfo{author}{\bibfnamefont{H.}~\bibnamefont{Liskien}} \bibnamefont{and}
  \bibinfo{author}{\bibnamefont{A.Paulsen}}, \bibinfo{journal}{Nuclear Data
  Tables (Nuclear Data Sect.A)} \textbf{\bibinfo{volume}{11}},
  \bibinfo{pages}{569} (\bibinfo{year}{1973}).

\bibitem[{\citenamefont{N.Jarmie}(1977)}]{ja77}
\bibinfo{author}{\bibfnamefont{J.}~\bibnamefont{N.Jarmie}},
  \bibinfo{journal}{Phys. Rev. C} \textbf{\bibinfo{volume}{16}},
  \bibinfo{pages}{15} (\bibinfo{year}{1977}).

\bibitem[{\citenamefont{Brown and Jarmie}(1990)}]{brown}
\bibinfo{author}{\bibfnamefont{R.~E.} \bibnamefont{Brown}} \bibnamefont{and}
  \bibinfo{author}{\bibfnamefont{N.}~\bibnamefont{Jarmie}},
  \bibinfo{journal}{Phys. Rev. C} \textbf{\bibinfo{volume}{41}},
  \bibinfo{pages}{1391} (\bibinfo{year}{1990}).

\bibitem[{\citenamefont{Ziegler}()}]{srim}
\bibinfo{author}{\bibfnamefont{J.}~\bibnamefont{Ziegler}},
  \urlprefix\url{http://www.srim.org/}.

\bibitem[{\citenamefont{Fuchs et~al.}(2007)\citenamefont{Fuchs, Sentoku,
  d'Humieres, Cowan, Cobble, Audebert, Kemp, Nikroo, Antici, Brambrink
  et~al.}}]{fuchs}
\bibinfo{author}{\bibfnamefont{J.}~\bibnamefont{Fuchs}},
  \bibinfo{author}{\bibfnamefont{Y.}~\bibnamefont{Sentoku}},
  \bibinfo{author}{\bibfnamefont{E.}~\bibnamefont{d'Humieres}},
  \bibinfo{author}{\bibfnamefont{T.~E.} \bibnamefont{Cowan}},
  \bibinfo{author}{\bibfnamefont{J.}~\bibnamefont{Cobble}},
  \bibinfo{author}{\bibfnamefont{P.}~\bibnamefont{Audebert}},
  \bibinfo{author}{\bibfnamefont{A.}~\bibnamefont{Kemp}},
  \bibinfo{author}{\bibfnamefont{A.}~\bibnamefont{Nikroo}},
  \bibinfo{author}{\bibfnamefont{P.}~\bibnamefont{Antici}},
  \bibinfo{author}{\bibfnamefont{E.}~\bibnamefont{Brambrink}},
  \bibnamefont{et~al.}, \bibinfo{journal}{Physics of plasmas}
  \textbf{\bibinfo{volume}{14}}, \bibinfo{pages}{053105}
  (\bibinfo{year}{2007}).

\bibitem[{\citenamefont{Clayton}(1983)}]{clayton}
\bibinfo{author}{\bibfnamefont{D.~D.} \bibnamefont{Clayton}},
  \emph{\bibinfo{title}{Principles of Stellar Evolution and Nucleosynthesis}}
  (\bibinfo{publisher}{University of Chicago Press}, \bibinfo{year}{1983}).

\bibitem[{\citenamefont{Forslund et~al.}(1977)\citenamefont{Forslund, Kindel,
  and Lee}}]{forslund}
\bibinfo{author}{\bibfnamefont{D.~W.} \bibnamefont{Forslund}},
  \bibinfo{author}{\bibfnamefont{J.~M.} \bibnamefont{Kindel}},
  \bibnamefont{and} \bibinfo{author}{\bibfnamefont{K.}~\bibnamefont{Lee}},
  \bibinfo{journal}{Phys. Rev. Lett.} \textbf{\bibinfo{volume}{39}},
  \bibinfo{pages}{284} (\bibinfo{year}{1977}).

\bibitem[{\citenamefont{Begay and Forslund}(1982)}]{forslund2}
\bibinfo{author}{\bibfnamefont{F.}~\bibnamefont{Begay}} \bibnamefont{and}
  \bibinfo{author}{\bibfnamefont{D.~W.} \bibnamefont{Forslund}},
  \bibinfo{journal}{Phys. Fluids} \textbf{\bibinfo{volume}{25}},
  \bibinfo{pages}{1675} (\bibinfo{year}{1982}).

\bibitem[{\citenamefont{Angulo et~al.}(1999)\citenamefont{Angulo, Arnould,
  Rayet, Descouvemont, Baye, Leclercq-Willain, Coc, Barhoumi, Aguer, Rolfs
  et~al.}}]{nacre}
\bibinfo{author}{\bibfnamefont{C.}~\bibnamefont{Angulo}},
  \bibinfo{author}{\bibfnamefont{M.}~\bibnamefont{Arnould}},
  \bibinfo{author}{\bibfnamefont{M.}~\bibnamefont{Rayet}},
  \bibinfo{author}{\bibfnamefont{P.}~\bibnamefont{Descouvemont}},
  \bibinfo{author}{\bibfnamefont{D.}~\bibnamefont{Baye}},
  \bibinfo{author}{\bibfnamefont{C.}~\bibnamefont{Leclercq-Willain}},
  \bibinfo{author}{\bibfnamefont{A.}~\bibnamefont{Coc}},
  \bibinfo{author}{\bibfnamefont{S.}~\bibnamefont{Barhoumi}},
  \bibinfo{author}{\bibfnamefont{P.}~\bibnamefont{Aguer}},
  \bibinfo{author}{\bibfnamefont{C.}~\bibnamefont{Rolfs}},
  \bibnamefont{et~al.}, \bibinfo{journal}{Nucl. Phys. A}
  \textbf{\bibinfo{volume}{656}}, \bibinfo{pages}{3} (\bibinfo{year}{1999}).

\bibitem[{\citenamefont{Kimura and Bonasera}()}]{kb-t}
\bibinfo{author}{\bibfnamefont{S.}~\bibnamefont{Kimura}} \bibnamefont{and}
  \bibinfo{author}{\bibfnamefont{A.}~\bibnamefont{Bonasera}}, \eprint{to be
  submitted}.

\bibitem[{\citenamefont{Awes et~al.}(1981)\citenamefont{Awes, Poggi, Gelbke,
  Back, Glagola, Breuer, and Viola}}]{PhysRevC.24.89}
\bibinfo{author}{\bibfnamefont{T.~C.} \bibnamefont{Awes}},
  \bibinfo{author}{\bibfnamefont{G.}~\bibnamefont{Poggi}},
  \bibinfo{author}{\bibfnamefont{C.~K.} \bibnamefont{Gelbke}},
  \bibinfo{author}{\bibfnamefont{B.~B.} \bibnamefont{Back}},
  \bibinfo{author}{\bibfnamefont{B.~G.} \bibnamefont{Glagola}},
  \bibinfo{author}{\bibfnamefont{H.}~\bibnamefont{Breuer}}, \bibnamefont{and}
  \bibinfo{author}{\bibfnamefont{V.~E.} \bibnamefont{Viola}},
  \bibinfo{journal}{Phys. Rev. C} \textbf{\bibinfo{volume}{24}},
  \bibinfo{pages}{89} (\bibinfo{year}{1981}).

\bibitem[{exf()}]{exfor}
\urlprefix\url{http://www-nds.iaea.org/exfor/exfor.htm}.

\end{thebibliography}
%



\end{document}